\documentclass{ws-p8-50x6-00}

\begin{document}

\title{Black Brane World Scenarios
\thanks{This talk is based on works in collaboration with W.S. Bae, Y.M. Cho,
Y. Kim and S.-J. Rey.}}
\author{Sei-Hoon Moon}
\address {School of Physics, Seoul National University, Seoul 151-742, Korea
\\E-mail: jeollo@phya.snu.ac.kr}

\maketitle

\abstracts{We consider a brane world residing in the interior region inside 
the horizon of extreme black branes. In this picture, the size of the horizon 
can be interpreted as the compactification size. The large mass hierarchy is 
simply translated into the large horizon size, which is provided by the 
magnitude of charges carried by the black branes. Hence, the macroscopic 
compactification size is a quantity calculable from the microscopic 
theory which has only one physical scale, and its stabilization is guaranteed 
from the charge conservation.}

In this talk, we will pay attention to the interior region bounded by a 
degenerated horizon of extreme black branes. The interior region possesses all 
of the features needed for large extra dimension (ADD)\cite{ADD} as well as 
Randall-Sundrum (RS) scenarios\cite{RS} provided that the 
interior region is regular. Since the interior region has finite volume while 
it asymptotes to the infinitely long AdS throat, the central region inside 
the near horizon region acts like a domain wall embedded in anti-de Sitter 
space ($AdS_5$) like the RS brane. The massless 4D graviton is clearly 
localized to the central region, reproducing the correct 4D Newtonian gravity. 

The size of the horizon can be interpreted as the compactification size in 
that the four-dimensional Planck scale $M_{pl}$ is determined by the 
fundamental scale $M_*$ of the higher-dimensional theory and the size of the 
horizon $r_H$ via the familiar relation $M_{pl}^2\sim M_*^{2+d}r_H^d$ and the 
effective gravity on a 3-brane residing in the central region has a transition 
from the four-dimensional to the higher-dimensional gravity around distances 
of the size of the horizon. This scenario gives a natural explanation
of the large mass hierarchy between the four-dimensional Planck scale $M_{pl}$
and the weak scale $m_{EW}$. The large mass hierarchy is translated into the
large size of the horizon. The large size of the horizon, {\it i.e.,} the small 
compactification scale is provided with the large magnitude of charge carried
by the black branes, that is, winding number or R-R charge. 
Hence, the macroscopic compactification size now is a quantity calculated from
the microscopic theory and its stabilization is guaranteed from the charge
conservation.

We will consider two types of black brane solutions. Firstly, we will show a
global black brane solution\cite{KMR,Moon} which is a black hole like defect 
solution to a scalar theory with global $O(d)$ symmetry coupled to higher 
dimensional gravity and is a $p$-dimensional extended object surrounded by a 
degenerated horizon. This solution is perfectly regular everywhere. 
Secondly, we will consider the supergravity solution of the D3-brane. 
Its interior region interpolates between the singularity and the near horizon 
region. When the singularity is smoothed out by stringy effects, the graviton
is localized in the central region. Further, even in the existence the gravity 
could be trapped, under the assumption of the unitary boundary condition for 
the graviton states\cite{BCM}.

\vspace{0.3cm}

The scalar theory with global $O(d)$ symmetry coupled to $D$-dimensional 
gravity, of which potential has minimum on the $(d-1)$-sphere with radius 
$v^2$ allows a black hole like extended defect solution, which we will call
as `global black $p$-brane'. The global black $p$-brane is a $p$-dimensional 
extended object surrounded by a degenerated horizon. This object is very 
similar to the black $p$-brane solutions of supergravity and string theories. 
However, this solution is perfectly regular everywhere, that is, either 
inside or outside of the horizon. 

Here, we will not write down the Einstein's equations of motion and the scalar
field equation of motion, and we will skip the procedure to find the global
black brane solution. These are treated in detail in Refs.[8,9].
We will begin just with a brief description for the spacetime of the global 
black brane. We introduce the following Schwartzschild-type metric ansatz:  
\begin{equation}\label{bbm}
ds^{ 2}=e^{2N(r)}B(r) \bar{g}_{\mu\nu}(x)dx^\mu dx^\nu+\frac{d^{2}r}{B(r)}
+r^{2}d\Omega^{2}_{d-1},
\end{equation}
where $\bar{g}_{\mu\nu}(x)$ is a general Ricci-flat metric on the brane, which 
satisfies $(p+1)$D Einstein equations $\bar{R}_{\mu\nu}(\bar{g})=0$.  
We have looked into the behavior of solutions at a few regions because it seems
almost impossible to find exact analytic solutions. 
Outside the core, the behavior of metric functions is controlled by the ratio 
between the scalar field energy density ($8\pi G_D/r^2$) and the cosmological 
constant ($|\Lambda|$). 
In the far region where the cosmological constant dominates over the field 
energy density. The metric (\ref{bbm}) has the form of 
\begin{eqnarray}\label{bbas}
ds^2\approx B_{\infty}r^2~\bar{g}_{\mu\nu}(x)dx^\mu dx^\nu
+\frac{dr^2}{B_{\infty}r^2}+r^2d\Omega_{d-1}^2, 
\end{eqnarray}
where $B_{\infty}\equiv 2|\Lambda|/(p+d)(p+d-1)$.
This corresponds to $D$-dimensional anti-de Sitter space ($AdS_D$) and can be 
changed to the form found in Refs.[3,4] using 
the proper radial distance $\chi(\equiv \int^r dr'/\sqrt{B(r')})$ as
\begin{eqnarray}\label{amet}
ds^2\approx e^{2\sqrt{B_\infty}\chi}~\bar{g}_{\mu\nu}(x)dx^\mu dx^\nu+d\chi^2
             +e^{2\sqrt{B_\infty}\chi}~d\Omega_{d-1}^2.
\end{eqnarray} 
The horizon occurs in a region where the field energy density and the 
cosmological constant are comparable, and the spacetime is approximated by 
\begin{equation}\label{bbhm}
ds^2\approx B_H[\sigma(r-r_H)]^{2(1-\alpha)}\bar{g}_{\mu\nu}(x)dx^\mu dx^\nu+
            \frac{dr^2}{B_H(r-r_H)^2}+r_H^2 d\Omega_{d-1}^2,
\end{equation}
where $\sigma$ is $-1$ for the interior region ($r<r_H$) and $+1$ for the
exterior region ($r>r_H$) and the curvature scale of this spacetime and the 
horizon size are given by 
\begin{eqnarray}
k^2&\equiv& B_H(1-\alpha)^2=\frac{2|\Lambda|}{(p+1)(p+d-1)},\label{kf1}\\
r_H^2&=&\frac{(p+d-1)(8\pi G_D-d+2)}{2|\Lambda|}. \label{rhf1}
\end{eqnarray}
It is easy to see relation between the near horizon solution and the cigar-like 
warped spacetime obtained in Refs.[3,4]. 
Introducing a new radial coordinate $\chi$($>0$) defined by 
$\exp(-k\chi)\equiv \sqrt{B_H}~[\sigma(r-r_H)]^{1-\alpha}$,
the metric Eq.(\ref{bbhm}) is rewritten as
\begin{equation}\label{grg}
ds^2\approx \exp\left(-2k\chi\right)\bar{g}_{\mu\nu}(x)dx^\mu dx^\nu
            +d\chi^2+r_H^2 d\Omega_{d-1}^2.
\end{equation}
Hence, among solutions obtained in Refs.[3,4], the meaningful 
solution with respect to the Randall-Sundrum scenario can naturally be 
interpreted as the near horizon geometry of a black $p$-brane. The two 
solutions (\ref{bbas}) and (\ref{grg}), which seemed to be disjointed in 
Refs.[3,4], can be matched to each other.
On the other hand, in the region between the brane core and the near horizon 
where the cosmological constant is negligible compared to the field energy 
density (if $\Lambda\ll M_*$, this region overwhelm the near horizon region), 
the shape of spacetime is largely dependent on dimension of transverse space. 
When $d=2$, the geometry of this region will resemble the Cohen-Kaplan 
spacetime\cite{CK}. When $d\geq3$, the solution in this region will be close 
to the global monopole metric \cite{Vil}. While we have not cleared up whether 
the solutions obtained in separated regions can be connected mutually or not, 
numerical works explicitly show that such black brane solutions exist.   

The interior region is of interest, because
it possesses all of the features needed for RS scenario, that is, it asymptotes 
to AdS space and its volume is finite and concentrated near the brane.

\vspace{0.3cm}
In the RS scenario, the necessary condition to have the usual Newtonian 
gravity on the brane is the existence of a massless 4D graviton 
localized on the brane. From now on, we will consider only $p=3$.
The existence of the massless graviton state is 
evident because Einstein's equations of motion always allows solutions 
with a general Ricci flat metric $\bar{g}_{\mu\nu}(x)$ and the massless 
4D graviton is simply the usual gravitational wave solution of 
linearized 4D vacuum Einstein equation. And the boundness of the 
massless graviton is equivalent to the condition that the 4D Planck 
scale $M_{pl}$ is finite. 
Examination of the 4D effective action yields 
the brane world Planck scale:
\begin{eqnarray}\label{pla}
M_{pl}^2= M_*^{d+2}\int dz^d \sqrt{g_D}~g^{00} 
            \sim M_*^{2+d}r_H^d .  
\end{eqnarray}
Thus, it is now clear that a massless state of the 4D graviton is bounded. 
The relation (\ref{pla}) tells that the 4D Planck scale is determined from 
the fundamental scale $M_*$ and the horizon size $r_H$ via the
familiar relation from usual Kaluza-Klein theories.
This implies that the horizon size $r_H$ can be interpreted as the effective 
size of $d$ compact extra dimensions, even though the interior region 
infinitely extends. From the point of view of one who lives in infinitely 
extended higher dimensional spacetime, this seems indeed a correct 
interpretation, clearly the interior region takes only a finite part with 
volume $\sim r_H^d$ of the infinite transverse space and the apparent infinite 
extent of the interior region is simply a result of the warping of the finite 
region of extra space by the gravity of the brane itself. This interpretation 
could be cleared up through a complete analysis of the effective 4D 
gravity.

In order to see the effective gravity on the brane, we introduce the 
perturbations by replacing $\bar{g}_{\mu\nu}(x)$ with 
$\eta_{\mu\nu}+h_{\mu\nu}(x,z)$ in Eq.(\ref{bbm}). Imposing the RS gauge on
$h_{\mu\nu}$, we can easily find the linearized field equation for 
$h_{\mu\nu}$. Making a change of variables  
$\xi=\int^r\sqrt{-g^{00}(r')g_{rr}(r')}~dr'$, 
$ \tilde{h}_{\mu\nu}= K~h_{\mu\nu}$ 
and separating
$\tilde{h}(\xi,\Omega)=\epsilon_{\mu\nu}e^{ip\cdot x}R_{m\ell}(\xi)
Y_\ell(\Omega)$, the linearized field equation can always be written into the 
form of an analog non-relativistic Schr\"{o}dinger equation: 
\begin{eqnarray}
&&\left[-\frac{\partial^2}{\partial\xi^2}+V_{eff}(\xi)\right]R_{m\ell}(\xi)
=m^2 R_{m\ell}(\xi),\label{sch} \\
{\rm with}~~&& V_{eff}(\xi)\equiv\frac{K''(\xi)}{K(\xi)}
+\ell(\ell+d-2)\frac{g_{00}(\xi)}{r(\xi)^2}, \label{efp}
\end{eqnarray}
where $K\equiv r^{(d-1)/2}g_{00}^{3/4}$ and $Y_\ell(\Omega)$ is a 
$d$-dimensional spherical harmonics. 
Here $\epsilon_{\mu\nu}$ is a constant polarization tensor and 
$m~(=\sqrt{p\cdot p})$ is the mass of continuum modes. 
The zero-mode wave function, with $m=\ell=0$, is easily identified as 
$R_{00}(\xi)=K(\xi)$. All of the important physics follows from a qualitative
analysis of the effective potential.

The Newtonian potential generated by a point source of mass $m^*$ localized 
on the brane then is 
\begin{eqnarray}\label{ngr}
U(|\vec{x}|) = -G_4\frac{m^*}{|\vec{x}|}
-\frac{m^*}{M_*^{d+2}}\sum_{\ell}
\int_{m\neq0}dm~m^{\delta}|R_{m\ell}(0)|^2~
\frac{e^{-m|\vec{x}|}}{|\vec{x}|}.
\end{eqnarray}
$m^{\delta}$ is contribution from more than one extra dimensions for measures 
of relevant density of states. If the extra dimensions are noncompactified, 
then $\delta=d-1$, simply. The factor of $m^{d-1}$ is just the $d$-dimensional 
plane wave continuum density of states (up to a constant angular factor).
In our case, $d-1$ extra dimensions are compactified to $S^{d-1}$ with radius
$r_H$. The modes with small $\ell$ behave as plane waves only in the 
radial direction, while the modes with large $\ell$ do in the full extra 
dimensions. Thus, $\delta$ will depend on $\ell$, that is, $\delta=0$ for modes
with small $\ell$ but $\delta = d-1$ for sufficiently large $\ell$.

In the original warped bulk model of Randall-Sundrum type scenario, the bulk 
cosmological constant $\Lambda$ had always been identified with the fundamental
scale for naive naturalness reasons. 
However, given our ignorance regarding the cosmological constant problem, we
do not feel any strong prejudice forcing $\Lambda$ to be of the order of the
fundamental scale $M_*^2$. We will simply treat $\Lambda$ as a parameter; we
know only that $r_H$ must be smaller than $\sim 1$mm from the present-day
gravity measurements if it should be interpreted as a compactification size.  
We will consider two limiting values of $\Lambda$; $|\Lambda|\sim M_*^2\sim 
M_{pl}^2$ and $|\Lambda|\sim10^{6-60/d}{\rm GeV}^2\ll M_*^2\sim m_{EW}^2$. 
We will call the first limit as Randall-Sundrum (RS) limit and the second 
limit as large extra dimension (ADD) limit.

\vspace{0.3cm}

In the RS limit, the interior region is well approximated by the near horizon
geometry Eq.(\ref{grg}) because both the core radius 
$r_c\sim(\sqrt{\lambda}v)^{-1/2}$ and the horizon size $r_H$ are of the order 
of the fundamental scale and so $8\pi G_D/r^2$ and $|\Lambda|$ are comparable 
in whole interior region. In this regime, since both curvature scales of 
$AdS_5$ and $S^{d-1}$ are of the order of the fundamental scale, the extra 
space effectively reduces to one-dimensional space because one who lives in 
the bulk could not observe the extra $S^{d-1}$ through low-energy processes 
under the fundamental scale. Then the 3-brane looks like a one-sided RS-brane 
embedded in a $AdS_5$ bulk spacetime. Hence the physics on 
the brane will be nearly the same with that of RS scenario. A difference 
arises due to massive KK modes with $\ell\neq0$ living in the $AdS_5$. 
However, this correction is strongly suppressed due to the repulsive 
centrifugal potential. Thus, the low-energy physics on the brane are 
imperceptibly different from those in the RS scenario.

\vspace{0.3cm}

In the ADD limit, the phenomenologically acceptable size of the horizon 
(cosmological constant) is $r_H\sim 10^{30/d-17}$cm 
($|\Lambda|\sim 10^{6-60/d}{\rm GeV}^2$) with 
$M_*\sim M_{EW}\sim10^3$GeV and $M_{pl}\sim10^{18}$GeV. 
Then the AdS region takes extremely tiny portion of the transverse space and 
the intermediate region, in which $8\pi G_D/r^2\gg|\Lambda|$,
between the core and the AdS region occupy most volume of transverse space.

The geometry of the intermediate region can be approximated by that of 
Cohen-Kaplan solution \cite{CK} when $d=2$ and those of global monopole 
spacetime \cite{Vil} when $d\geq3$. It is easy to explicitly calculate the 
potential Eq.(\ref{efp}) using the Cohen-Kaplan and the global monopole metric. 
The explicit calculations show that, when $d=2,3$, the potential is attractive 
or zero in the intermediate region. However, it is not needed to specify 
details of the the shape of the central region of the potential, because 
the potential is not repulsive in the intermediate region and the central 
region is localized within the the AdS length scale $k^{-1}$.
Then the leading corrections to the long-range gravitational potential are 
in fact identical to those in the RS limit, as shown in Ref.[7]. 
The short distance gravity is dominated by continuum modes with $m\geq k$ and
$\ell\gg1$ and $(4+d)$D gravity appears at distances $|\vec{x}|\ll r_H$.
Hence, in this regime, the brane world gravity behaves like that of RS limit 
except that it experience a transition from $\sim1/|\vec{x}|$ to 
$\sim 1/|\vec{x}|^{d+1}$ around the distances of the size of the horizon, which
is now much bigger than the weak scale, {\it i.e.,} $r_H\gg m_{EW}$.
Hence, the behavior of the gravity on the brane is exactly what we would expect
by interpreting the size of the horizon as a \lq\lq compactification radius".

When $d\geq4$, the physics will be quite different from that of when $d=2,3$ 
due to the repulsive potential in the intermediate region. 
Since the hight of the potential is $V_{eff}(\sim r_c)\sim M_{EW}^2$,
the continuum modes with mass $m\ll M_{EW}$ have suppressed wave functions 
near the core and so, for distances $|\vec{x}|\gg M_{EW}^{-1}$, 
the gravity behaves as in RS limit and is nearly four-dimensional. 
On the other hand, for distances $|\vec{x}|\ll m_{EW}$, the gravity is 
$D$-dimensional. This regime is phenomenologically interesting because the 
physics on the brane seems very distinct from both the RS and the ADD 
scenarios, in that the gravity is maintained to be four-dimensional down to
distances of $m_{EW}^{-1}$, despite introducing the large extra dimensions
to obtain the 4D Planck scale. Furthermore, there are no light moduli fields
associated with the large size of $S^{d-1}$ because they are strongly 
suppressed near the 3-brane.
 
\vspace{0.3cm}
In the ADD limit, we have obtained not only the phenomenology of theories with
large extra dimensions for $d=2,3$ but also somewhat distinct one for $d\geq4$.
In both cases, the hierarchy problem could be resolved via the large horizon 
size $r_H$ in the sense of the conventional large extra dimension scenarios.
However, there still remains a hierarchy between $M_{EW}$ and $|\Lambda|$,
because such large size of the horizon is provided by only the tiny bulk 
cosmological constant. This hierarchy could be stable in the sense that small 
changes of $|\Lambda|$ have small effects to the physics on the brane as 
expected in ADD scenarios. An unpleasant point is the fact that $|\Lambda|$ is 
a Lagrangian parameter that is not free from the radiative correction due to 
various bulk fields and needs to be stabilized by a symmetry, such as 
supersymmetry. In absence of a clear stabilization mechanism, it however seems 
natural $|\Lambda|$ to be of the order of the fundamental scale. 
From this aspect, the Randall-Sundrum limit only seems to be natural.

At this point, an immediate question is whether the such large size of the 
horizon (or equivalently the small compactification scale) can be calculated 
in the theory that has only one physical scale of the order of the weak scale. 
The answer seems to be positive. Usually the size of the horizon of an extreme
black brane is determined by the magnitude of charge (equivalently mass)
carried by the black brane. It does not seem uncommon to have solutions
with huge magnitude of charge; the world around us abounds with solutions 
that has much larger charge than the electron's. 

Up to now, we have considered only brane solutions with unit winding 
number. This was because the conditions of spherical symmetry and regularity
at the origin allows only solutions with unit winding number when $d\geq3$. 
While cylindrically symmetric solutions with arbitrarily large winding 
number $n$ are allowed when $d=2$. The solutions with winding number $n$ can 
easily be obtained simply replacing $8\pi G_D$ with $8\pi G_Dn^2$ in the
solution with unit winding number for $d=2$. 
Then the radius of the horizon is given from Eq.(\ref{rhf1}) in terms of 
unrescaled parameters by $r_H^2=16\pi v^2n^2/M_*^4|\Lambda|$.
We assume all scales to be in the same order with the fundamental scale, 
that is, intrinsically there exist only one physical scale. 
Then the horizon size is given by $r_H\sim n~M_*^{-1}$ simply.  
Putting $M_{*}\sim m_{\rm EW}\sim 10^3$GeV and demanding that $r_H$ and $n$ 
are chosen to reproduce the observed four-dimensional Planck scale 
$M_{pl}\sim 10^{18}$GeV yields $n\sim 10^{15}$. 
Hence, the global black brane solution with large winding number of 
$n\sim 10^{15}$ seems to provide a dynamical determination of the hierarchy 
between the four-dimensional Planck scale and the weak scale without 
requiring any additional small scale from the theory that has intrinsically 
only one physical scale. Furthermore, the such large compactification size is 
now stabilized via the charge conservation law.
 
\vspace{0.5cm}

The interior region of the D3-brane interpolates between the singularity and
the AdS throat. The similar analysis of the graviton states to that for the 
global black brane shows that even in the existence of the singularity the 
massless graviton can be localized in the central region of the interior 
region under the unitary boundary condition\cite{CK,grem}.
However, since we do not have a detailed understanding of the singularity,
we will not be able to make any rigorous claim whether such boundary 
conditions are what we want. Thus, we will only assume that the singularity is 
smoothed out by the true short-distance theory of gravity, namely, string 
theory. Essentially the source for the R-R field strength is sitting at the 
singularity. Since the only known source for R-R 5-form field strength is 
D3-brane, we are naturally expected to see a stack of D3-branes when we probe
the singularity with energy over the string scale.  

The analog Schr\"{o}dinger potential resembles with that for the global black 
brane with $d=2$. And so the gravity on a 3-brane residing in the central 
region 
behaves like what we would expect by interpreting the horizon size $r_H$ as a
compactification radius. Since the effective Planck scale is determined via
the relation $M_{pl}^2=r_H^6/(24\pi^3g_s^2l_s^8)$ and the size of the horizon is
given by $r_H^4=4\pi l_s^4g_sN$, the hierarchy between the string scale 
$l_s^{-1}$ and the four-dimensional Planck scale $M_{pl}$ can be provided 
with the R-R charge of the amount of 
$N=\left(2\pi^{3/2}l_s^2g_s^{1/2}M_{pl}^2\right)^{2/3}$. 
If we naively assume that the string scale is of the order of the weak scale, 
{\it i.e.,} $l_s\sim m_{EW}$, then the amount of the R-R charge needed
to generate the hierarchy is $N\sim 10^{22}$ and the size of the horizon
is $\sim 10^{-12}$cm.

Perhaps the deepest consequence of the above picture is that it gives a natural
explanation of the large mass hierarchy and gives a good reason of why string
theory would necessarily choose such type of compactification geometry.
The large mass hierarchy is translated into the large size of the horizon. 
The large size of the horizon, {\it i.e.,} the large compactification size 
can be calculated in string theory that has only one physical scale $l_s$ and 
is determined by the amount of the R-R charge $N$ carried by the D3-brane. 
Furthermore, the stabilization of such large compactification size is strictly 
supported by the conserved D3-brane charges. 

We don't need to compactify whole spacetime introducing the compactification
manifold. The geometry needed for the Randall-Sundrum type brane world can be
obtained simply from the noncompact ten-dimensional spacetime by means of 
formation of a large cluster of D3-branes. 

\vspace{0.5cm}

Let us conclude with some discussions. The above picture may provide new 
perspectives on problems associated to the brane world scenarios. 
Both non-Ricci flat metric $\hat{g}_{\mu\nu}(x)$ and metric dependent on the
extra dimensional coordinates $z^i$, {\it i.e.,} $\hat{g}_{\mu\nu}(x,z)\sim 
\eta_{\mu\nu}+ h_{\mu\nu}(x,z)$, correspond to excitations upon the extremal 
black brane background, which is the ground state of the black $p$-brane. 
That is, in the existence of such excitations the black brane becomes now 
non-extremal and its Hawking temperature is not zero. Thus, such excitations
will be diluted through Hawking radiation. First, the continuum modes are one 
of such excitations. The analysis on the singular behavior of the continuum 
modes at the horizon\cite{CG} should be reexamined because it has done on the 
rigid AdS background, while in the existence of such excitation we lose 
the AdS background. Second, the Poincar\'{e} invariance in the longitudinal 
direction will persist even in the presence of the quantum corrections to the
brane tension because the quantum corrections also will be diluted, so that
no 4D cosmological constant generated.
Third, this picture provides a new physical mechanism that can solve the 
cosmological flatness problem. Even though our world brane was highly bent
initially, the bending should have been diluted as the black brane Hawking 
radiates and evolves toward the extremal one. Since the entropy density of 
our universe is minutely small, our world brane seems to be 
embedded in the interior region of very near-extremal black brane.  
Finally, the another flatness problem\cite{CKR} associated with the 
approximate Lorentz invariance in the longitudinal direction also can be 
resolved because the bulk curvature will be diluted via Hawking process.

\vspace{0.5cm}
We wish to acknowledge discussions with H.M. Lee, J.D. Park and S. Yi.

\end{document}